\documentclass{aa}  
\usepackage{graphicx}
\usepackage{natbib}
\usepackage{txfonts}
\bibpunct{(}{)}{;}{a}{}{,}
\begin{document} 

\title{An Updated Census of RR~Lyrae Stars in the Globular Cluster $\omega$~Centauri (NGC\,5139)\thanks{Based on observations collected at the European Organisation for Astronomical Research in the Southern Hemisphere, Chile, with the VISTA telescope (project ID 087.D-0472, PI R. Angeloni).}} 
   \author{C. Navarrete\inst{\ref{inst1}, \ref{inst2}} \and
          R. Contreras Ramos\inst{\ref{inst1}, \ref{inst2}} \and
          M. Catelan\inst{\ref{inst1}, \ref{inst2}, \ref{inst3}} \and
          C. M. Clement\inst{\ref{inst4}} \and
          F. Gran\inst{\ref{inst1}, \ref{inst2}} \and
          J. Alonso-Garc\'ia\inst{\ref{inst1}, \ref{inst2}} \and\\
          R. Angeloni\inst{\ref{inst1}, \ref{inst3}, \ref{inst5}, \ref{inst6}} \and
          M. Hempel\inst{\ref{inst1}, \ref{inst2}} \and
          I. D\'ek\'any\inst{\ref{inst2}, \ref{inst1}} \and
		  D. Minniti\inst{\ref{inst7}, \ref{inst2}, \ref{inst1}}
          }         
% % 
   \institute{Instituto de Astrof\'isica, Pontificia Universidad Cat\'olica de Chile,
              Av. Vicu\~na Mackenna 4860, 782-0436 Macul, Santiago, Chile\\
              \email{cnavarre;rcontre;mcatelan@astro.puc.cl} \label{inst1} \and
              Millennium Institute of Astrophysics, Santiago, Chile\label{inst2} \and
              Centro de Astro-Ingenier\'ia, Pontificia Universidad Cat\'olica de Chile,
              Santiago, Chile\label{inst3} \and
			  Department of Astronomy and Astrophysics, University of Toronto,
              Toronto, Ontario, M5S3H4, Canada \label{inst4} \and
              Max-Planck-Institut f\"ur Astronomie, Konigstuhl 17, 69117, 
              Heidelberg, Germany \label{inst5} \and 
			  Gemini Observatory, Colina El Pino, Casilla 603, La Serena, Chile \label{inst6} \and 
			  Departamento de Ciencias F\'isicas, Universidad Andr\'es Bello, Fern\'andez Concha 700, 759-1538 Santiago, Chile \label{inst7}
			  \\
             }

   \date{Received Aug 20, 2014; accepted Jan 7, 2015}

  \abstract
  {}
   {$\omega$~Centauri (NGC~5139) contains large numbers of variable stars of different types and, 
   in particular, more than a hundred RR~Lyrae stars. A homogeneous (in terms of instrument, image 
   quality, and time coverage) sample of high-quality near-infrared (NIR) 
   RR~Lyrae light curves could thus be gathered 
   by performing an extensive time-series campaign aimed at this object. 
   We have conducted a variability survey of $\omega$~Cen in the   
   NIR, using ESO's 4.1m Visible and Infrared Survey Telescope for Astronomy (VISTA). This is the   
   first paper of a series describing our results.}
   %Accurate coordinates, magnitudes and 
   %periods are mandatory in order to derive such database.
   {$\omega$~Cen was observed using VIRCAM mounted on VISTA. 
   A total of 42 and 100 epochs in $J$ and $K_{\rm S}$, respectively, were obtained, distributed over 
   a total timespan of 352~days. Point-spread function
   photometry was performed using DAOPHOT and DoPhot in the inner and outer regions of the cluster, 
   respectively. Periods of the known variable stars were improved when necessary using an  
   ANOVA analysis.}
   {An unprecedented homogeneous and complete NIR catalogue of RR~Lyrae stars in the field of 
   $\omega$~Cen was collected, allowing us to study, for the first time, all the RR~Lyrae stars 
   associated to the cluster, except 4 located far away from the cluster center. Membership status, 
   subclassifications between RRab and RRc subtypes, periods, amplitudes, and mean magnitudes 
   were derived for all the stars in our sample. Additionally, 4 new RR~Lyrae stars were discovered, 
   2 of them with high probability of being cluster members. The distribution of $\omega$~Cen stars 
   in the Bailey (period-amplitude) diagram is also discussed. Reference lines in this plane, for 
   both Oosterhoff type I (OoI) and Oosterhoff type II (OoII) components, are provided, both in 
   $J$ and $K_{\rm S}$. 
   }
   {In the present paper, we clarify the status 
   of many (candidate) RR~Lyrae stars that had been unclear in previous studies. This includes 
   stars with anomalous positions in the color-magnitude diagram, uncertain periods or/and variability
   types, and possible field interlopers. We conclude that $\omega$~Cen hosts a total of 88 RRab and 
   101 RRc stars, for a grand total of 189 likely members. We confirm that most RRab stars in 
   the cluster appear to belong to an OoII component, as previously found using visual data.} 
%    Our full, updated catalogue of variable stars in 
%    $\omega$~Cen, including the complete list of RR~Lyrae stars, type II Cepheids, SX~Phoenicis stars, 
%    eclipsing binaries, and anomalous Cepheids, along with their NIR properties, will be provided 
%    in the next paper in this series.}
   
   \keywords{Stars: variables: RR~Lyrae~-- Stars: variables: general~-- Stars: horizontal-branch~-- 
             Galaxy: globular clusters: individual: $\omega$~Centauri (NGC~5139)~--
             Infrared: stars~-- Surveys~-- Catalogs
               }

   \maketitle
%
%________________________________________________________________

\section{Introduction}\label{Intro}

  RR~Lyrae stars (RRLs) are well known for being reliable distance indicators and tracers of 
  the old stellar populations of galaxies \citep[][and references therein]{S95, C09, B11}. In the 
  dawning era of time-series infrared (IR) photometric surveys, new observations in the near- and 
  mid-IR have recently been providing large and homogeneous samples of RRLs in different environments 
  \citep[e.g.,][]{D13, K14}, which can be used to improve our understanding of this variability 
  class and as tracers of structure in the nearby Universe.
  
  Especially noteworthy, in this context, is the fact that RRLs follow well-defined period-luminosity 
  (PL) relations in the IR \citep[e.g.,][]{L90, MC04}. Furthermore, the interstellar extinction 
  in $K_{\rm S}$ is approximately a tenth of the one in the optical, thus making it possible to 
  observe RRLs even in the most obscured regions of our Galaxy, e.g., its inner bulge. Despite these 
  advantages, there are still few extensive variability studies in the near-infrared (NIR),  
  due to a number of factors, including the lack, until recently, of large-format detectors, which rendered  
  NIR surveys extremely costly, in terms of observing time. In addition, the low amplitudes of many 
  types of variables, including RRLs, in the NIR, compared to the optical, have hindered the 
  detection of new variables using these filters. The situation has only recently started to change, 
  particularly with the start of operations of ESO's 4.1m Visible and Infrared Survey Telescope 
  for Astronomy (VISTA), located in Cerro Paranal, Chile \citep{jeea06,E10}. 
  
  The VISTA Variables in the V\'ia L\'actea (VVV) ESO Public Survey is an extensive variability 
  study in the $K_{\rm S}$-band of the Galactic bulge and an adjacent portion of the disk \citep[see][and references 
  therein]{M10, C11, C13}. In order to automatically classify the large samples of variable stars detected 
  by this survey, our team started to build a database of well-defined, high-quality NIR light 
  curves for various variability classes – the VVV Templates Project\footnote{\tt http://www.vvvtemplates.org} 
  (\citealt{C13}; \citealt{raea14}). In this context, $\omega$~Centauri (NGC~5139) was identified 
  early on as one of the most suitable targets to obtain template light curves, given that it hosts 
  a large number and variety of variable stars.
  
  $\omega$~Cen stands out as one of the three most RRL-rich globular clusters known, only 
  surpassed by M3 \citep[NGC~5272, with 230 RRLs;][]{C04} and M62 \citep[NGC~6626, with 224 
  RRLs;][]{rcea05,CR10}. From the several variability studies that have been carried out in the 
  optical over the years with different telescopes and temporal baselines, a large (although still 
  incomplete) catalogue of RRLs in $\omega$~Cen has been produced (\citealt{C01}; C. M. Clement's 
  online catalogue,\footnote{\tt http://www.astro.utoronto.ca/$\sim$cclement/read.html} 2013 
  edition, hereinafter C13). A wide-field, NIR variability survey of $\omega$~Cen allows us to 
  derive and calibrate the NIR PL relations of several different types of stars, including 
  SX~Phoenicis, type II Cepheids, and RRL stars. In this sense, our study of $\omega$~Cen with 
  VIRCAM@VISTA has provided the largest homogeneous NIR sample of RRLs ever collected 
  in any globular cluster, leading to tight PL relations, as described in our next papers in 
  this series on $\omega$~Cen variables. 
  
  $\omega$~Cen is the most massive and luminous globular cluster in the Milky Way, and it may 
  have been even more massive in the past \citep[e.g.,][and references therein]{vc11}. Nowadays, 
  $\omega$~Cen is well known for hosting at least two stellar populations with metallicity peaks 
  centered at ${\rm [Fe/H]} \sim -1.6$ and $-1.1$~dex \citep[][and references therein]{B04, J13}. 
  In addition to presenting variations in [Fe/H] that may be correlated with age, it has been found 
  that $\omega$~Cen's stars also present variations in the abundances of many other elements, 
  both heavy and light, in addition to helium (at a level $\Delta Y \sim 0.17$~dex) and age 
  \citep[e.g.,][]{N04, P05, D11,vc11}.

  Despite the fact that He enhancement appears to be required to explain the lower  
  color-magnitude diagram (CMD) morphology of $\omega$~Cen, an impact on the cluster's variable   
  stars has not yet been established. \cite{S06} found that the RRL stars with metal-intermediate 
  composition (${\rm [Fe/H]}\sim-1.2$), fainter than the bulk of the dominant metal-poor population 
  (with primordial helium abundances and ${\rm [Fe/H]}\sim-1.7$~dex), are in good agreement with  
  the corresponding horizontal branch (HB) models with $Y = 0.246$. More recently, \cite{M11} studied  
  the impact of the helium content on the RRL properties based on evolutionary and pulsation  
  models, and again could not find evidence of He enhancement among the RRL stars. The most  
  likely interpretation of these results is that any He-enhanced populations in the cluster  
  likely end up far to the blue on the zero-age HB, thus being unlikely to populate the classical
  instability strip \citep[e.g.,][]{J13}.
 
  In this paper, in addition to describing our VISTA observations, we also update the census of 
  RRLs in $\omega$~Cen. In this sense, we provide, using our data and comparing with previous 
  studies, a comprehensive assessment of their number, types, membership probabilities, and light 
  curves, and also improved light curves for several RRL variables. We point out some erroneous 
  cross-identifications between the \cite{W07} and \cite{K04} catalogues, and propose a change  
  in classification for a few stars. The membership status of all the suspected field stars is 
  revisited. Last but not least, we present four new RRLs, two of which we classify as genuine 
  cluster members. In future papers of this series, we will study the NIR PL relations for RRL, 
  type II Cepheids, and SX~Phe stars, and present the full updated catalogue of variable stars (including 
  also eclipsing binaries) in the $\omega$~Cen field. 

% ________________________________________________________________

 \section{Observations}\label{Observations}
 
  A total of 42 and 100 epochs in $J$ and $K_{\rm S}$, respectively, were taken using VIRCAM 
  \citep{D06} mounted on VISTA \citep{jeea06,E10}. All of the images were centered on the cluster 
  center. The heart of the VIRCAM camera is a $4\times 4$ array of Raytheon 
  VIRGO IR detectors ($2048\times 2048$ pixels), with a pixel size of $0\farcs34$ \citep{D06}. The 
  size of a uniformly covered field (also called a ``tile'') is 1.501~deg$^2$, thus allowing 
  unprecedented spatial coverage in the NIR bandpasses. The effective field of view (FoV) of 
  the camera is larger than the previous studies of \cite{K04} and \cite{W07}, both carried out 
  in the visible, but not large enough to cover $\omega$~Cen in its whole extension (its tidal 
  radius is $\simeq 52 \arcmin$; e.g., \citealt{ffea06}, and references therein). Our observations 
  have a baseline of 352 days, which provides an unprecedented phase coverage in the NIR. 
  
  The characteristics of the observations, data reduction and photometry extraction are the same as 
  those explained in \cite{N13}. In particular, the images were reduced by the Cambridge Astronomy 
  Survey Unit (CASU),\footnote{\tt http://casu.ast.cam.ac.uk/surveys-projects/vista} and crowded-field, 
  point-spread function (PSF) photometry was performed using the DAOPHOT~II/ALLFRAME \citep{S87,pbs94} 
  and DoPhot \citep{S93,jagea12} photometry packages for the cluster central (i.e., the innermost 
  $\sim 10\arcmin$) and outer regions, respectively. 

  The DoPhot and ALLFRAME photometries were calibrated independently, using in both cases the 
  aperture photometry provided by CASU, which is in the VISTA photometric system. As described in 
  detail in \citet{M10}, the CASU pipeline determines the photometric zero points using 2MASS 
  stars \citep{msea06} in the pawprint frames, then transforming these 2MASS magnitudes into VISTA 
  magnitudes according to the equations derived by \cite{H09}. Magnitudes for common stars derived 
  using both packages were then compared, finding an excellent agreement in the magnitude calibration 
  (at the level of RRL stars, the average $K_{\rm S}$-band magnitude difference between the DoPhot 
  and ALLFRAME magnitudes is $\sim 0.01$~mag only).  
  
%________________________________________________________________

\section{Results}\label{sec:results} 
 
  Time-series photometry of most of the previously known variable stars listed by C13 (from V1 to 
  V410) and \cite{W07} was recovered using a cross-match between those catalogues and our complete 
  PSF photometric catalogue (i.e., photometry for all the stars at all the observed epochs). The 
  matching radius was set at 1\arcsec. When one star was not recovered under this procedure, the 
  \cite{S09} catalogue was used as reference. In the \citeauthor{S09} catalogue, variable stars 
  from V1 to V293 are listed, most of them with updated coordinates based on 2MASS observations. 
  The cross-match using the \citeauthor{S09} catalogues allowed us to recover 16 variable stars 
  (including 8 RRL, 3 SX~Phoenicis, 4 eclipsing binaries, and 1 spotted variable) that were not 
  recovered using the C13 catalogue. When neither of these reference catalogues gave a match for a 
  certain variable star, an increasing matching radius was used, until a known variable star was 
  found. To pinpoint the latter, the time series for all stars that were inside the matching radius 
  were phased using the star's period as listed by C13, until reasonable agreement was found. As a 
  consistency check, periods for those stars were also calculated using the ANOVA algorithm \citep{S89}, 
  in an effort to determine if the known variable is present in the area covered by the matching 
  radius but with a different period according to our observations. This approach led to the 
  discovery of V411 as a new RRL star \citep{N13}, which \cite{W07} had previously (but incorrectly) 
  claimed to be the same as V144 from the \cite{K04} catalogue. 

\begin{figure}[ht]
   \centering
   \includegraphics[width=0.495\textwidth]{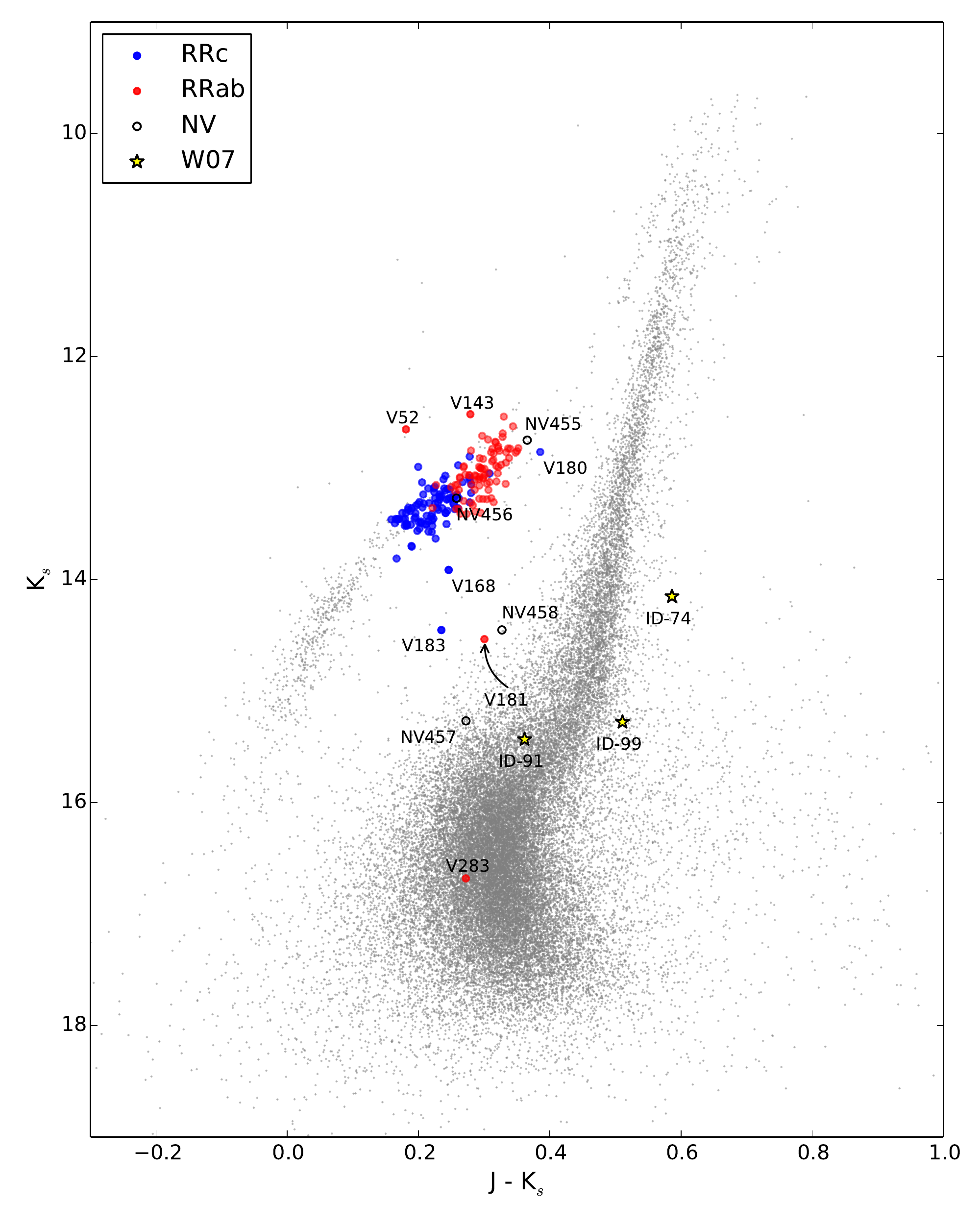}
   \includegraphics[width=0.495\textwidth]{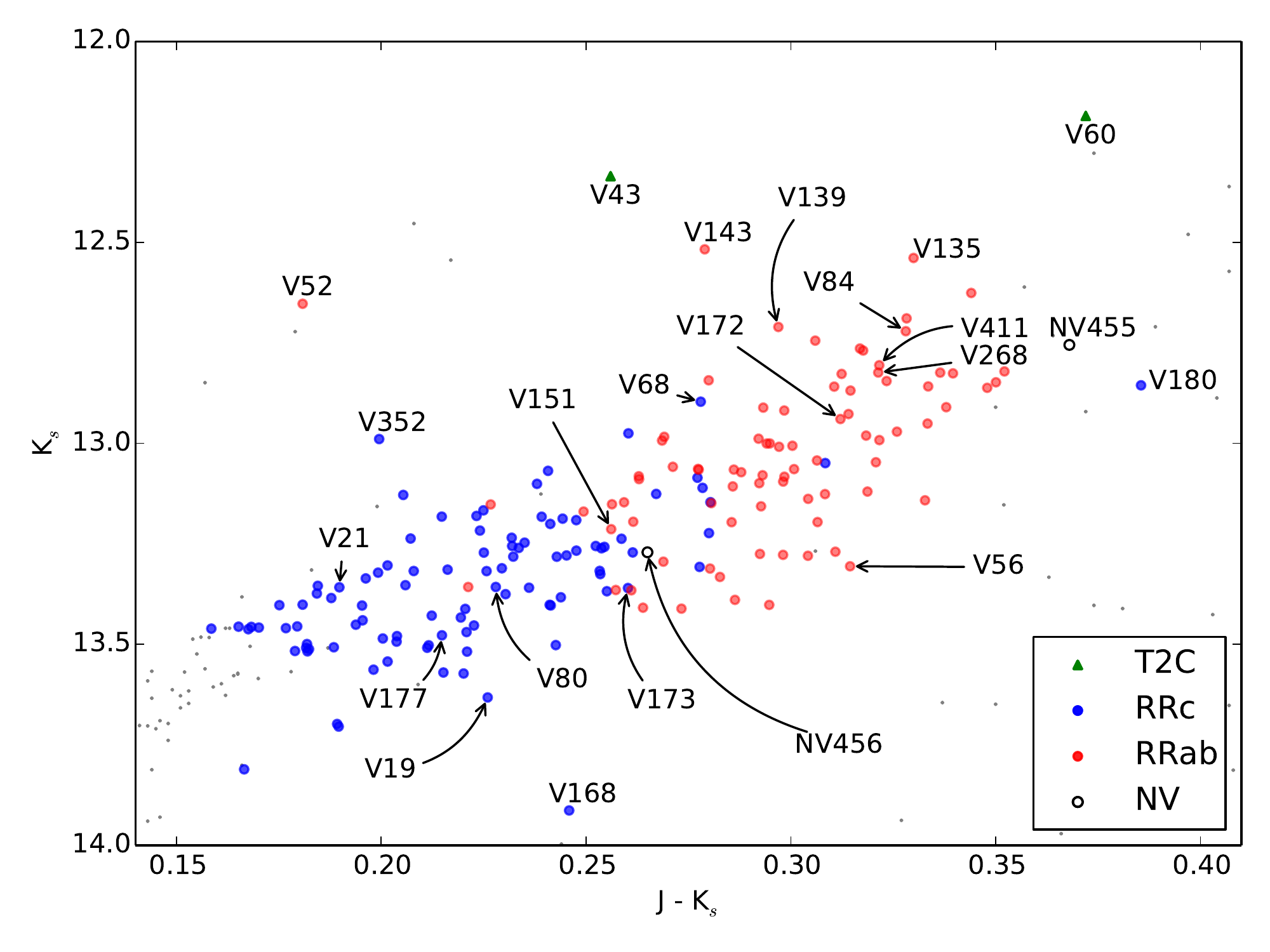}
   \caption{\textit{Top:} CMD of the central part of the cluster 
   (11~arcmin$^{2}$, grey dots). ab- and c-type RRL are marked as red and blue circles, 
   respectively. RRL from \cite{W07} are marked as star symbols. Open circles mark the new 
   RRLs found in this work, with NV455 and NV456 (labeled) likely being cluster members 
   (see Sect.~\ref{newv}). Field, blended, or erroneously classified stars are labeled 
   and explained in the text. \textit{Bottom:} Zoom-in around the RRL region in the CMD of $\omega$~Cen,
   with a few noteworthy stars labeled. V43 and V60, marked with green 
   triangles, are known type II Cepheids.}
   \label{CMD}
   \end{figure}

  Figure~\ref{CMD} shows the $\omega$~Cen $K_{\rm S}$, $(J-K_{\rm S})$ CMD, 
  as obtained from our ALLFRAME photometry for the central regions (see Sect.~\ref{Observations}). 
  In this diagram, RRLs from the whole observed area are superimposed as red and blue circles for 
  RRab (i.e., fundamental-mode) and RRc (first-overtone) types, respectively. Star symbols indicate 
  variables from \cite{W07} (see Sect.~\ref{Weldrake}), and new variables are shown as open 
  circles (see Sect.~\ref{newv}). Most of the RRLs are located in the horizontal branch (HB), as 
  expected for bona-fide cluster members, but a few stars fall close to the main sequence, or even 
  at the turn-off point. A zoom-in around the RRL instability strip (IS) region is shown in the 
  bottom panel of Figure~\ref{CMD}, where there are also two type II Cepheids, marked as green 
  triangles. All the RRL stars labeled in both figures were considered in our analysis and studied 
  in terms of their periods, variability types, and cluster membership. In the following, we clarify 
  the status of many of the RRL stars in $\omega$~Cen that had been unclear in previous studies. 
  This includes stars with anomalous CMD positions, uncertain periods or/and variability types, 
  and possible field interlopers.

  Table~\ref{table:1} lists the characteristics of 21 previously known RRL that either have 
  anomalous positions in the CMD, were suggested as field stars by previous studies (and fall 
  within our FoV), or do not have periods (or even variability type) listed in the literature. 
  Column~1 lists the ID of the variables according to C13, except for the two stars from \cite{W07}, 
  because they are not included in the C13 catalogue (ID-91, ID-99); columns~2 and 3 show the 
  coordinates based on VISTA astrometry (which in turn is tied to the 2MASS system); columns~4 and 
  5 display the intensity-weighted mean $J$ and $K_{\rm S}$ magnitudes, respectively; column~6 
  gives our derived periods, while the variability type (RRab, RRc, eclipsing binary [EB], or 
  anomalous Cepheid [ACEP]) is given in column~7; the distance to the cluster center for each 
  variable is listed in column~8. The membership probability for the stars, according to the two 
  main proper motion studies of $\omega$~Cen, are listed in columns~9 and 10, as provided by 
  \citet[][hereafter vL00]{vL00} and \citet[][hereafter B09]{B09}, respectively.\footnote{\cite{vL00} 
  constructed a proper motion catalogue for stars in $\omega$~Cen based on observations of 
  photographic plates of the cluster, taken between 1931 and 1983 with the Yale-Columbia 66-cm 
  telescope. The covered area has a radius of $\sim 30\arcmin$ (about 60\% of the cluster's tidal 
  radius). The \cite{B09} proper motion catalogue was in turn derived from astrometry based on CCD 
  images taken with the Wide-Field Imager on the 2.2-m telescope at ESO La Silla, with a precision 
  of $\sim 7$~mas. The B09 observations were carried out with a baseline of 4 years, and extend to 
  a limiting magnitude of $B\sim 20$~mag, which is up to four magnitudes deeper than previous studies.} 
  Finally, column~11 shows the membership status according to our photometric study (m:~member, 
  f:~field). A brief description of each of these variables is presented in the Appendix.
 
   \begin{table*}
  \caption{Properties of member and field RRLs with previously uncertain status}             
  \label{table:1}      
  \centering          
  \begin{tabular}{ccccccccccc}
  \hline\hline
  ID    & RA          & DEC      & $\langle J\rangle$ & $\langle K_{\rm S}\rangle$ & $P$\tablefootmark{(a)} & Type & $d$\tablefootmark{(b)} & $\mathcal{P}_{\mu}^{\rm vL00}$ & $\mathcal{P}_{\mu}^{\rm B09}$ & Memb.\\
        & (J2000)     & (J2000) & (mag) & (mag) & (days) & & (arcmin) & (\%) & (\%) & status\\
  \hline
  V19   & 13:27:30.12 &  -47:28:05.74  & 13.852 & 13.631 & 0.299551 & RRc        &  7.27 & 100 & 100 & m \\
  V21   & 13:26:11.15 &  -47:25:59.30  & 13.538 & 13.362 & 0.380812 & RRc        &  6.71 & 100 & 99  & m \\
  V52   & 13:26:35.15 &  -47:28:04.33  & 12.831 & 12.634 & 0.660386 & RRab       &  2.16 & 53  & 45  & m \\
  V56   & 13:25:55.44 &  -47:37:44.44  & 13.612 & 13.301 & 0.568023 & RRab       & 12.53 & 98  & 100 & m \\
  V68   & 13:26:12.79 &  -47:19:36.12  & 13.168 & 12.897 & 0.534696 & RRc/ACEP?  & 10.87 & 100 & 100 & m?\\
  V80\tablefootmark{(c)}
        & 13:28:54.96 &  -47:30:16.42  & 13.568 & 13.360 & 0.37718  & RRc        & 21.62 & out & out & m \\
  V84   & 13:24:47.47 &  -47:49:56.18  & 13.031 & 12.712 & 0.579873 & RRab/ACEP? & 29.24 & 0   & out & m? \\
  V143  & 13:26:42.59 &  -47:27:28.98  & 12.783 & 12.510 & 0.820734 & RRab       &  1.51 & 90  & 99  & m \\
  V151  & 13:28:25.31 &  -47:16:00.01  & 13.455 & 13.211 & 0.407756 & RRc        & 20.94 & out & out & m \\
  V168  & 13:25:52.74 &  -47:32:03.19  & 14.164 & 13.916 & 0.321299 & RRc        & 9.78  & 0   & 0   & f \\
  V172\tablefootmark{(d)}
        & 13:27:55.04 &  -47:04:38.50  & 13.220 & 12.927 & 0.7380   & RRab       & 26.72 & out & out & m \\
  V173  & 13:29:43.13 &  -47:16:53.80  & 13.624 & 13.360 & 0.3590   & RRc        & 32.05 & out & out & m \\      
  V177\tablefootmark{(e)}
        & 13:29:04.15 &  -47:36:21.42  & 13.690 & 13.479 & 0.31469  & RRc        & 24.31 & out & out & m \\
  V180  & 13:28:15.22 &  -47:40:32.02  & 13.201 & 12.839 & 0.39006  & W~UMa      & 18.93 & 0   & out & f \\
  V181  & 13:30:00.35 &  -47:48:44.96  & 14.823 & 14.535 & 0.5884   & RRab       & 38.17 & out & out & f \\
  V183  & 13:29:39.50 &  -47:30:18.36  & 14.690 & 14.449 & 0.29603  & RRc        & 29.13 & out & out & f \\
  V268  & 13:26:35.11 &  -47:26:11.15  & 13.156 & 12.827 & 0.812922 & RRab       &  3.30 & 70  & 99  & m \\
  V283  & 13:27:36.53 &  -47:46:40.44  & 16.947 & 16.682 & 0.517233 & RRab       & 19.73 & out & out & f \\      
  V411\tablefootmark{(f)}  
        & 13:26:40.77 &  -47:28:17.00  & 13.123 & 12.796 & 0.8449   & RRab       &  1.20 & 100 & 99  & m \\
  V433\tablefootmark{(g)}
        & 13:29:03.59 &  -47:48:58.25  & 14.737 & 14.151 & 0.6681   & ?          & 30.58 & out & out & f \\
  ID-91 & 13:26:14.03 &  -47:53:08.20  & 15.782 & 15.430 & 0.637238 & RRab?      & 25.00 & out & out & f \\
  ID-99 & 13:24:40.44 &  -47:45:22.54  & 15.777 & 15.275 & 0.670361 &  ?         & 27.07 & out & out & f \\
  \hline
  \end{tabular}
  \tablefoot{
  \tablefoottext{a}{Periods are from \cite{K04}, except for V80, V172, V173, V177, V180, V181, V183 and V411, which were derived by us.}
  \tablefoottext{b}{Distance to the cluster center was calculated using $\omega$~Cen's center as reported by SIMBAD: RA~= 13:26:47.28, DEC~= -47:28:46.1, both in J2000.}
  \tablefoottext{c}{ID-53 in \citet{W07}.}
  \tablefoottext{d}{ID-18 in \citet{W07}.}
  \tablefoottext{e}{ID-49 in \citet{W07}.}
  \tablefoottext{f}{V411 has the same coordinates and periods as first reported by \cite{N13}. Mean magnitudes 
  differ from that paper because the values presented in this paper, but not in \cite{N13}, are intensity-averaged  mean magnitudes.}
  \tablefoottext{g}{ID-74 in \citet{W07}.}
  }
  \end{table*}  
    
%________________________________________________________________

  \subsection{Bailey diagram}\label{sec:bailey}
  As shown by \citet{C00} based on visual data, individual stars in $\omega$~Cen predominantly follow 
  a locus on the Bailey diagram (period-amplitude plane) that is 
  characteristic of Oosterhoff type II (OoII) systems, with individual RRL typically having longer 
  periods at a given amplitude than their Oosterhoff type I (OoI) counterparts. The details, systematics, 
  astrophysical interpretation and importance of the Oosterhoff phenomenon are extensively discussed
  in \citet{C09}. 
  
\begin{figure}[ht!]
   \centering
   \includegraphics[width=\hsize]{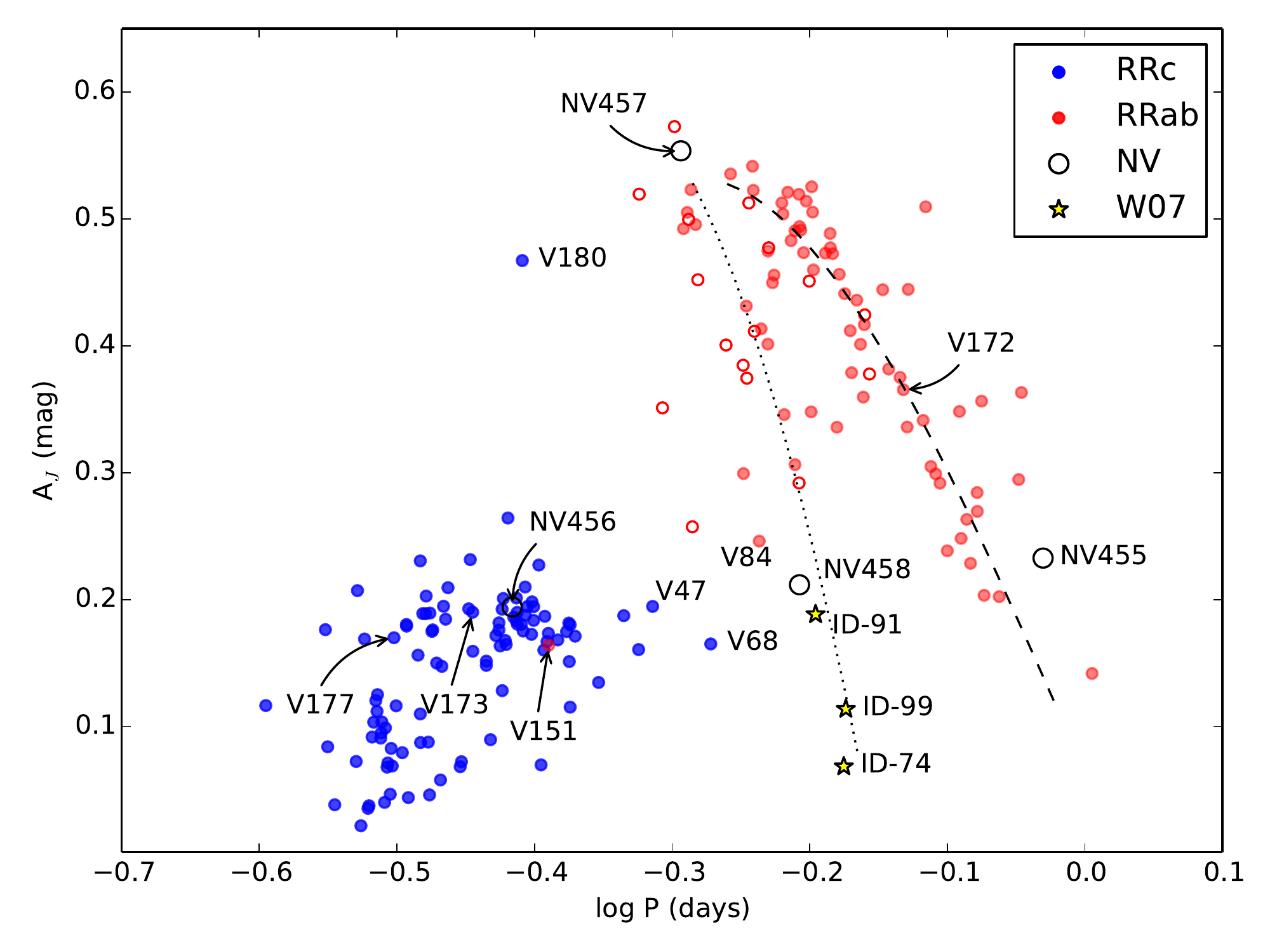}
   \includegraphics[width=\hsize]{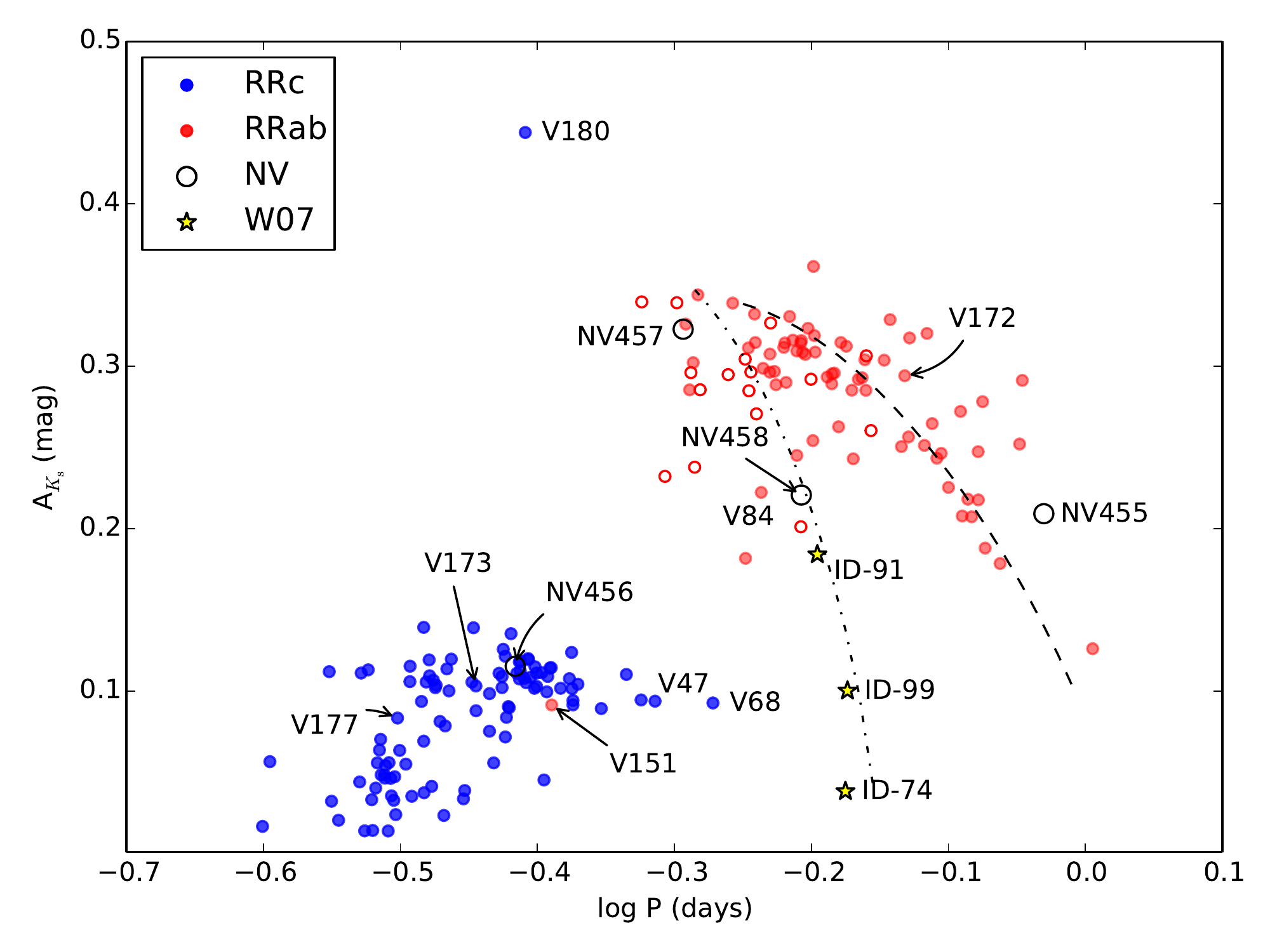}
   \caption{Bailey diagrams for all the RRL stars known in the $\omega$~Cen field, based on $J$- ({\em top}) 
   and $K_{\rm S}$-band {\em bottom} light curves. The overplotted lines define representative loci of 
   ``OoII-like'' ({\em dashed lines}) and ``OoI-like'' ({\em dotted lines}) behavior 
   for the ab-type RRL, obtained as described in the text. RRab stars with 
   documented Blazhko effect are marked with open circles. Clearly, V180 has too large an amplitude 
   for its claimed RRc status; instead, we believe that the star is a contact binary of the W~UMa  
   type with twice the alternative ``RRc-like'' period. V151 is clearly an RRc star. 
   }
   \label{bailey}
   \end{figure}

  Figure~\ref{bailey} shows, for the first time, the distribution of $\omega$~Cen RRL stars in the 
  NIR Bailey diagram, both in the $J$ (top panel) and in the $K_{\rm S}$ (bottom panel) bands. 
  RRab stars with documented Blazhko effect \citep{sb07} in the catalogue of \cite{K04} are 
  plotted as open red circles. Several features can be apprehended from these plots. First, the 
  positions occupied by c- and ab-type RRL stars are clearly very different, as also seen in the 
  visual. Second, there is significantly more scatter in the $K_{\rm S}$-band Bailey diagram than 
  is present in $J$. This is, at least in part, due to the smaller amplitudes in $K_{\rm S}$, 
  compared to $J$. Third, and as it is also commonly seen in the visible, the Blazhko stars tend to 
  have smaller amplitudes, at a given period, than their non-Blazhko counterparts. This is particularly 
  noteworthy as the actual amplitude modulation that is brought about by the Blazhko effect is not 
  clearly seen in our NIR data for any of $\omega$~Cen's RRab stars, which could be either because 
  the amplitude changes are too small in the NIR (notwithstanding the fact that Fig.~\ref{bailey}
  might hint at amplitude modulations exceeding 0.1~mag in at least some cases), and/or because our 
  time coverage is insufficient 
  to reveal long-term modulation (recall that our observations have a timespan of $\sim 300$~days, 
  with gaps). Indeed, to the best of our knowledge, there have been no clear-cut measurements of 
  the Blazhko effect in the IR, even though \citet{asea08} have reported increased 
  scatter in their NIR light curves of RR~Lyr itself, which they attribute to the star's well-known  
  Blazhko-type light curve modulation. 
  Fourth, there is a tendency for the locus occupied by ab-type RRL to flatten out, 
  as one moves from bluer to redder wavelengths. In fact, this trend appears to continue towards 
  the mid-IR regime, where the relation eventually becomes completely flat \citep{tgea14}. Fifth, 
  and again as previously emphasized by \citet{C00}, most RRab stars in $\omega$~Cen tend to clump 
  around a fairly well-defined locus, particularly in the $J$-band Bailey diagram. Based on 
  our $J$ amplitudes for the ``OoII-like'' RRab stars, this locus is fitted by the following expression: 
  
  \begin{equation}
  \mathcal{A}_{J}^{\rm OoII} = 0.064 - 2.481 \, \log P + 10.345 \, \left(\log P\right)^3, 
  \label{eq:ooii}
  \end{equation}  

  \noindent where the period is given in days. The amplitude-period correlation given by 
  Equation~\ref{eq:ooii} is plotted as a dashed line in the $J$-band Bailey diagram (Figure~\ref{bailey}). 
  That this indeed corresponds to \citeauthor{C00}'s OoII component can be seen by the fact that this 
  line is significantly displaced towards longer periods, compared to the OoI reference line (also shown 
  in Figure~\ref{bailey}). 
  
\begin{figure}
   \centering
   \includegraphics[width=\hsize]{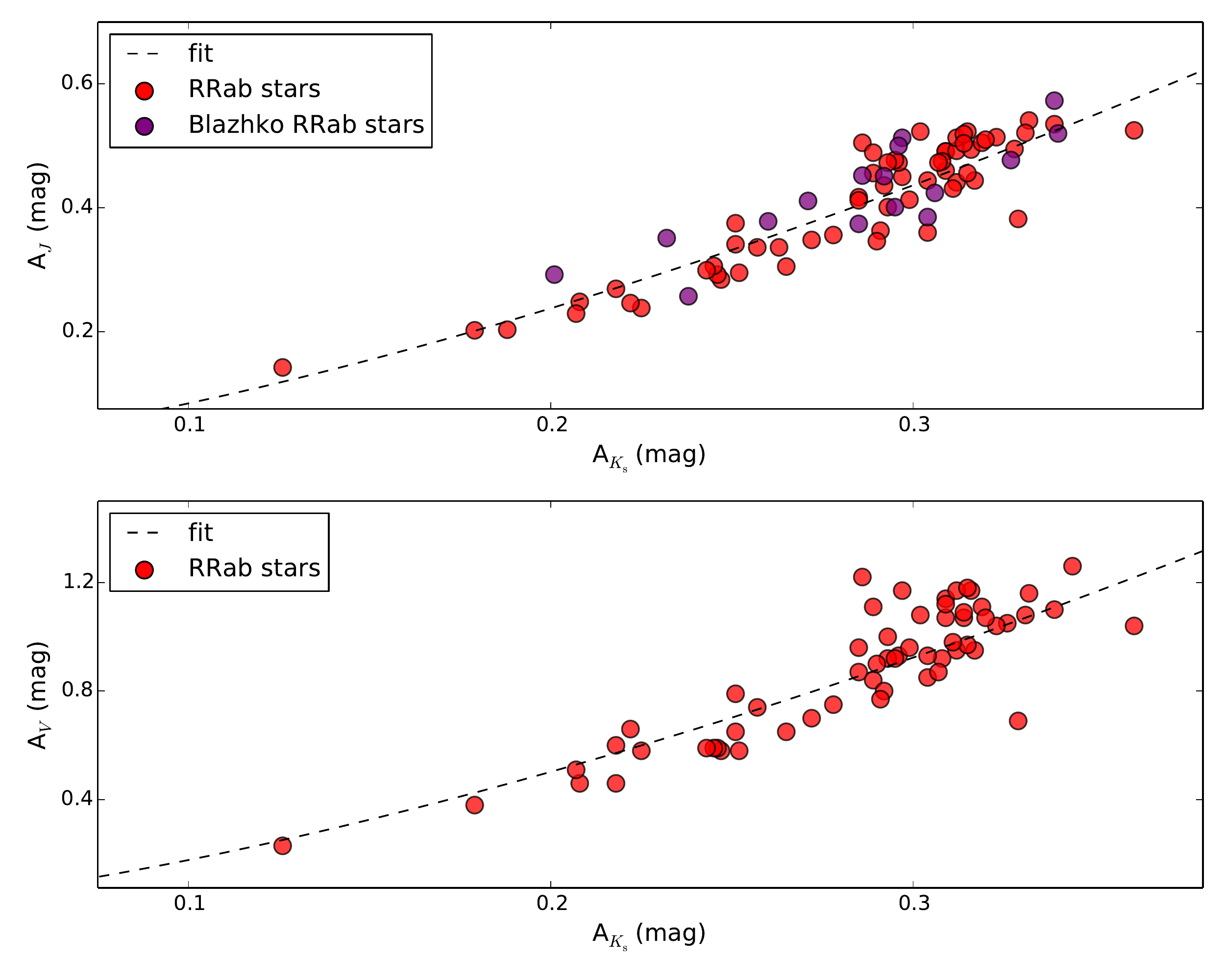}
   \caption{Relation between $K_{\rm S}$- and $J$-band amplitudes ({\em top}) and $K_{\rm S}$- and $V$-band amplitudes 
   ({\em bottom}) for $\omega$~Cen RRab stars. The dashed lines are the fit to the data (eq.~\ref{eq:aj-vs-ak} and \ref{eq:aksv}). 
   }
   \label{fig:aj-vs-ak}
   \end{figure}

  As can be seen from Figure~\ref{fig:aj-vs-ak}, there is a fairly tight relation 
  between $J$- and $K_{\rm S}$-band RRL amplitudes, which is described by the following 
  expression:    
  
  \begin{equation}
  \mathcal{A}_{J} = 2.6 \, \mathcal{A}^{3/2}_{K_{\rm S}}. 
  \label{eq:aj-vs-ak}
  \end{equation}  
  
  \noindent  Equation~\ref{eq:aj-vs-ak} was used to obtain the $K_{\rm S}$-band equivalent 
  of Equation~\ref{eq:ooii}. OoII reference loci obtained in this way are overplotted on both panels 
  of Figure~\ref{bailey} as dashed lines, clearly showing the expected period shift with respect to 
  the (minority) ``OoI-like'' RRL stars in $\omega$~Cen. 
    
  Unfortunately, there are few stars in $\omega$~Cen that belong to the OoI component \citep{C00}, 
  and thus we cannot define the OoI reference line reliably based solely on our data. Therefore, 
  to obtain an OoI reference locus in the NIR, we have used as a reference the $V$-band OoI 
  line from \citet{ccea05} and \citet{mzea10}, transformed into the NIR bandpasses by means 
  of the following relations: 

  \begin{equation}
  \mathcal{A}_{J} = 0.46 \, \mathcal{A}_{V}, 
  \label{eq:ajv}
  \end{equation}  

  \begin{equation}
  \mathcal{A}_{K_{\rm S}} = 0.32 \, \mathcal{A}_{V}^{2/3}, 
  \label{eq:aksv}
  \end{equation}  

  \noindent both of which are based on $\omega$~Cen RRab stars, with visual amplitudes coming from 
  \citet{K04}. The corresponding $J$- and $K_{\rm S}$-band OoI lines are shown in Figure~\ref{bailey} 
  as dotted lines. The presence of OoI-like RRL stars in $\omega$~Cen is confirmed, and is especially evident in the $J$-band Bailey diagram.

  These interrelations involving $\mathcal{A}_{V}$, $\mathcal{A}_{J}$, and 
  $\mathcal{A}_{K_{\rm S}}$, compared to those previously presented 
  by \citet{mfea08}, provide a description of the amplitudes that is physically better motivated, 
  as they more properly describe their asymptotic behavior towards vanishing light-curve amplitudes. 
    
   \begin{figure*}
   \centering
   \includegraphics[width=0.9\hsize]{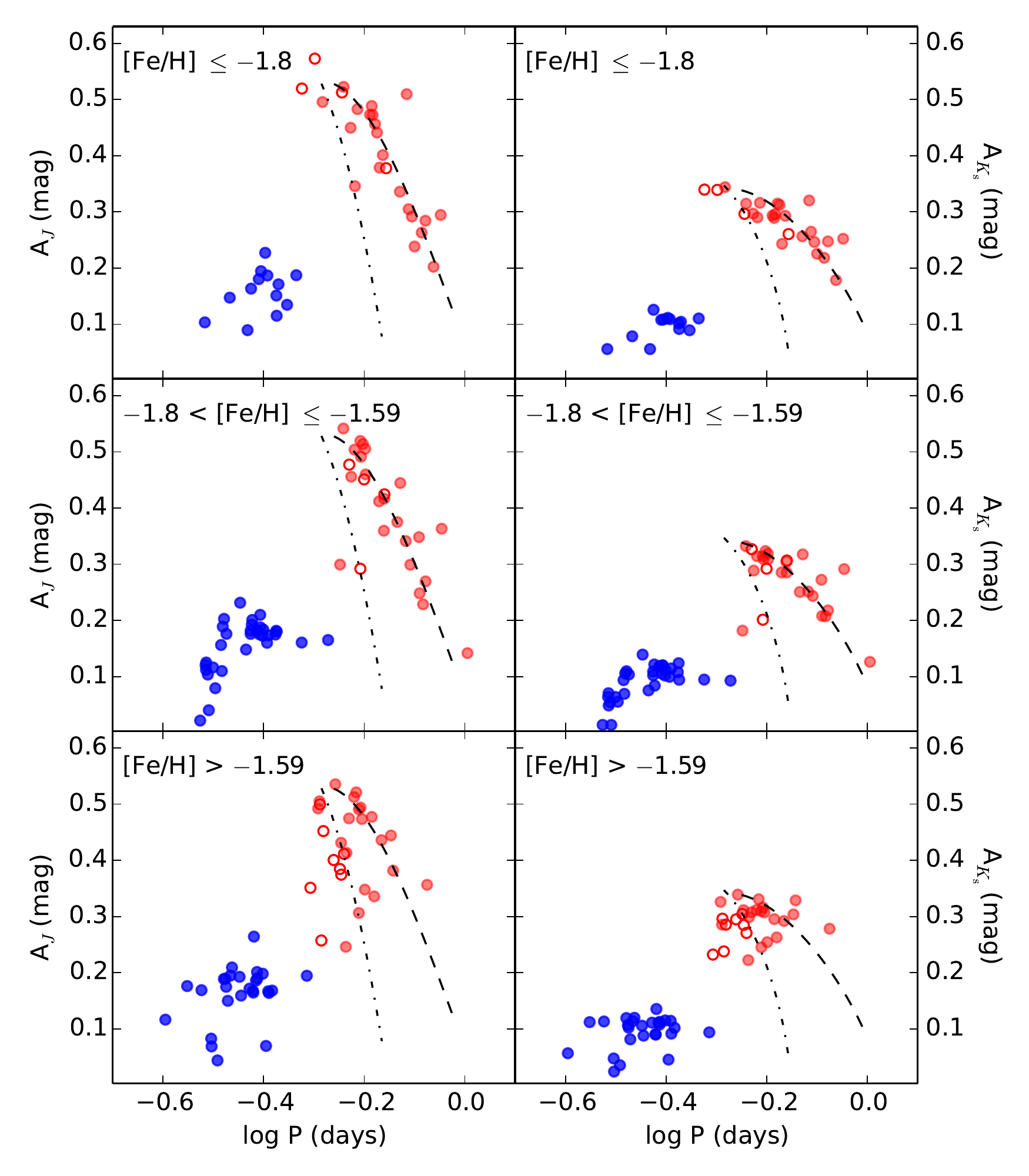}
   \caption{As in Figure~\ref{bailey}, but separating the RRL stars into three metallicity bins, 
   from lowest ({\em top}) to highest ({\em bottom}). The sizes of the bins were chosen so as to 
   include the same number of RRab stars in each. 
   }
   \label{fig:bailey_panels}
   \end{figure*}
 
  It is still a matter of debate whether the position of stars on the Bailey diagram, and therefore 
  the Oosterhoff type, is determined primarily by the metallicity of the cluster or by its HB morphology
  \citep[][and references therein]{rcea05,CR10,C00, R00}. 
  In this respect, $\omega$~Cen provides an interesting test of the relative importance of HB morphology 
  and [Fe/H] in defining the Oosterhoff classification, given that the cluster possesses a 
  predominantly blue HB, as in the case of most OoII clusters, but it also shows at least two 
  well-defined metallicity peaks, separated by $\approx 0.5$~dex. 
  To further investigate the possible impact of metallicity upon the distribution of RRL stars in 
  the NIR Bailey diagrams, Figure~\ref{fig:bailey_panels} again shows the Bailey diagrams in 
  both $J$ and $K_{\rm S}$, but separating the RRL into three metallicity bins, based on 
  [Fe/H] values from \cite{S06} and \cite{R00}. The bin sizes were chosen so as to ensure 
  an equal number of RRab stars in each bin.   
  For stars with [Fe/H] measurements in both catalogues, the 
  metallicities of \citeauthor{S06} were preferred. 
  
  As shown in Figure~\ref{fig:bailey_panels}, $\omega$~Cen can still 
  (not unexpectedly) be safely classified as predominantly an OoII GC, which is in perfect agreement 
  with its predominantly blue HB morphology. This figure also shows 
  that the NIR Bailey diagrams for $\omega$~Cen ab-type RRL stars is largely insensitive to metallicity, 
  in the sense that the average OoII and OoI loci, defined as above, adequately represent the mean loci 
  occupied by RRL in each individual metallicity bin also. It is interesting to note, however, that the OoI/OoII 
  number ratio does change with metallicity: in the lowest-metallicity bin, there are virtually no 
  RRL stars close to the OoI locus; conversely, in the highest-metallicity bin, RRab stars are more 
  evenly divided between OoII and OoI sub-classes. This largely confirms previous results from 
  \cite{R00} based on visual data, even though the impact of metallicity at low [Fe/H] was stronger 
  in their case. Thus, HB morphology and metallicity appear both to be important in determining 
  Oosterhoff type.

  %________________________________________________________________

  \begin{figure*}
   \centering
   \includegraphics[width=0.195\textwidth]{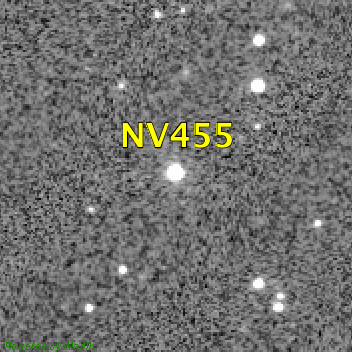}
   \includegraphics[width=0.195\textwidth]{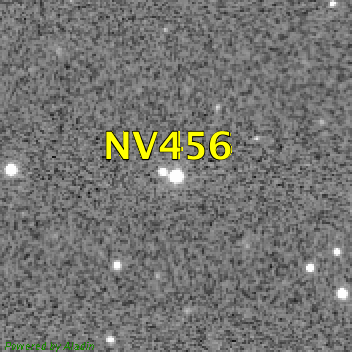}
   \includegraphics[width=0.195\textwidth]{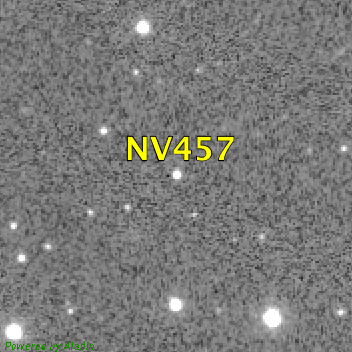}
   \includegraphics[width=0.195\textwidth]{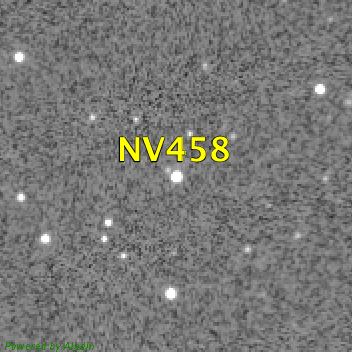}
   \caption{Finding charts for the 4 new RRL candidates that were discovered in our VISTA images. 
   North is up and East to the left, and each stamp covers a field of 1~arcmin$^{2}$.  
   The discovered candidates are the stars at the center of each stamp. 
   The first two stamps correspond to 2 new RRL members of $\omega$~Cen, based on their position in the 
   CMD (Fig.~\ref{CMD}), whereas the remaining ones, according to the same CMD, are field interlopers,
   in the cluster background.
   \label{fcharts}}
   \end{figure*}

  \begin{table*}[ht!]
  \caption{New RR~Lyrae candidate stars\tablefootmark{(a)}}             
  \label{table:2}      
  \centering          
  \begin{tabular}{ccccccccc}
  \hline\hline
   ID   & RA          & DEC          & $\langle J\rangle$ & $\langle K_{\rm S}\rangle$ &  $P$ &  Type & $d$        & Memb.\\                                                                                        
        & (J2000)     & (J2000)      & (mag)              & (mag)                & (days)  &       & (arcmin) & status \\  
  \hline
  NV455 & 13:27:53.94 & -46:55:43.93 &      13.132        &       12.760         & 0.932469 & RRab & 34.92    & m \\
  NV456 & 13:22:14.49 & -47:24:21.64 &      13.530        &       13.270         & 0.383516 & RRc  & 46.33    & m \\
%  NV457 & 13:30:15.83 & -47:51:59.00 &      14.907        &       14.524         & 0.204980 & RRc  & 42.09    & f \\
  NV457 & 13:29:54.56 & -47:50:46.00 &      15.530        &       15.265         & 0.508593 & RRab & 38.45    & f \\
  NV458 & 13:30:00.09 & -47:13:05.63 &      14.772        &       14.441         & 0.620313 & RRab & 36.23    & f \\
  \hline
  \end{tabular}
  \tablefoot{
  \tablefoottext{a}{Columns have the same meaning as in Table~\ref{table:1}.} 
  } 
  \end{table*}

%________________________________________________________________

  Interestingly, Figure~9 in \cite{R00} suggests that the $\omega$~Cen ab-type RR~Lyrae with the 
  longest periods (i.e., $P \gtrsim 0.71$~d) all belong to the metal-intermediate bin, but this is 
  not immediately obvious from our plot. We tested whether this difference is simply due to the 
  different metallicity bins selected by \citeauthor{R00} and our study. Indeed, according to our 
  analysis, even using the \cite{S06} metallicities, the long-period RRab's in $\omega$~Cen RRab's 
  do indeed occupy a very limited range in [Fe/H], between $-1.81$ and $-1.64$. This can be 
  understood in terms of the evolutionary scenario proposed by \cite{L91}, according to which 
  the HB component over this metallicity range is extremely blue, thus preferentially producing 
  overluminous, long-period RRL stars.

  \subsection{Low-amplitude RRc stars}\label{low}
  
  Some RRc stars in $\omega$~Cen are known to have very low amplitudes in the visible, at the level 
  of $\mathcal{A}_{V} < 0.1$~mag \citep{K04}. Specific examples include V281 
  ($\mathcal{A}_{V} \approx 0.07$~mag), V344 ($\mathcal{A}_{V} \approx 0.08$~mag), V357 
  ($\mathcal{A}_{V} \approx 0.05$~mag), and V399 ($\mathcal{A}_{V} \approx 0.05$~mag). 
  We do not expect to recover the variability of these stars from our NIR observations, 
  because the amplitudes in the NIR are even lower than in the optical \citep{S95}. 
  Indeed, none of these stars were found to be variable in our data. 
 
  In this context, we also note that clean light 
  curves for V339 and V340 could not be recovered either, even though their $V$ amplitudes  
  ($\mathcal{A}_{V} \approx 0.11$~mag and 0.17~mag, respectively) are perhaps high enough
  that one might have expected non-negligible $J$ and $K_{\rm S}$ amplitudes. A possible  
  explanation for their non detection in our data is their marked multiperiodic behavior,  
  which renders even their $V$-band light curves somewhat noisy \citep{K04}. 
  
  \subsection{New RR~Lyrae stars}\label{newv}
  
  To detect additional, previously unknown variable stars in and around $\omega$~Cen, 
  we performed a new search for variable stars over the complete area covered by 
  our observations. Our procedure consisted in considering as candidate variables all those stars 
  whose light curves implied root-mean-square magnitude deviations that were higher than the average trend
  for non-variable stars at the same magnitude. The search was independently performed in $J$ and 
  $K_{\rm S}$. With the goal of specifically detecting RRL and other periodic variables, 
  the four most prominent periods for each star, as returned by the ANOVA algorithm \citep{S89}, 
  were used to plot phase-folded variables. From this search, four variable 
  star candidates have periods and amplitudes in accordance with their being RRL stars. 
  Figure~\ref{fcharts} shows the finding chart for these new discovered RRL candidates. None 
  of them appears to present the Blazhko effect. 
  The remaining candidates, which may belong to different variability classes, will be discussed
  in future papers of this series. 
  
\begin{figure*}
   \begin{center}
   \includegraphics[width=0.75\hsize]{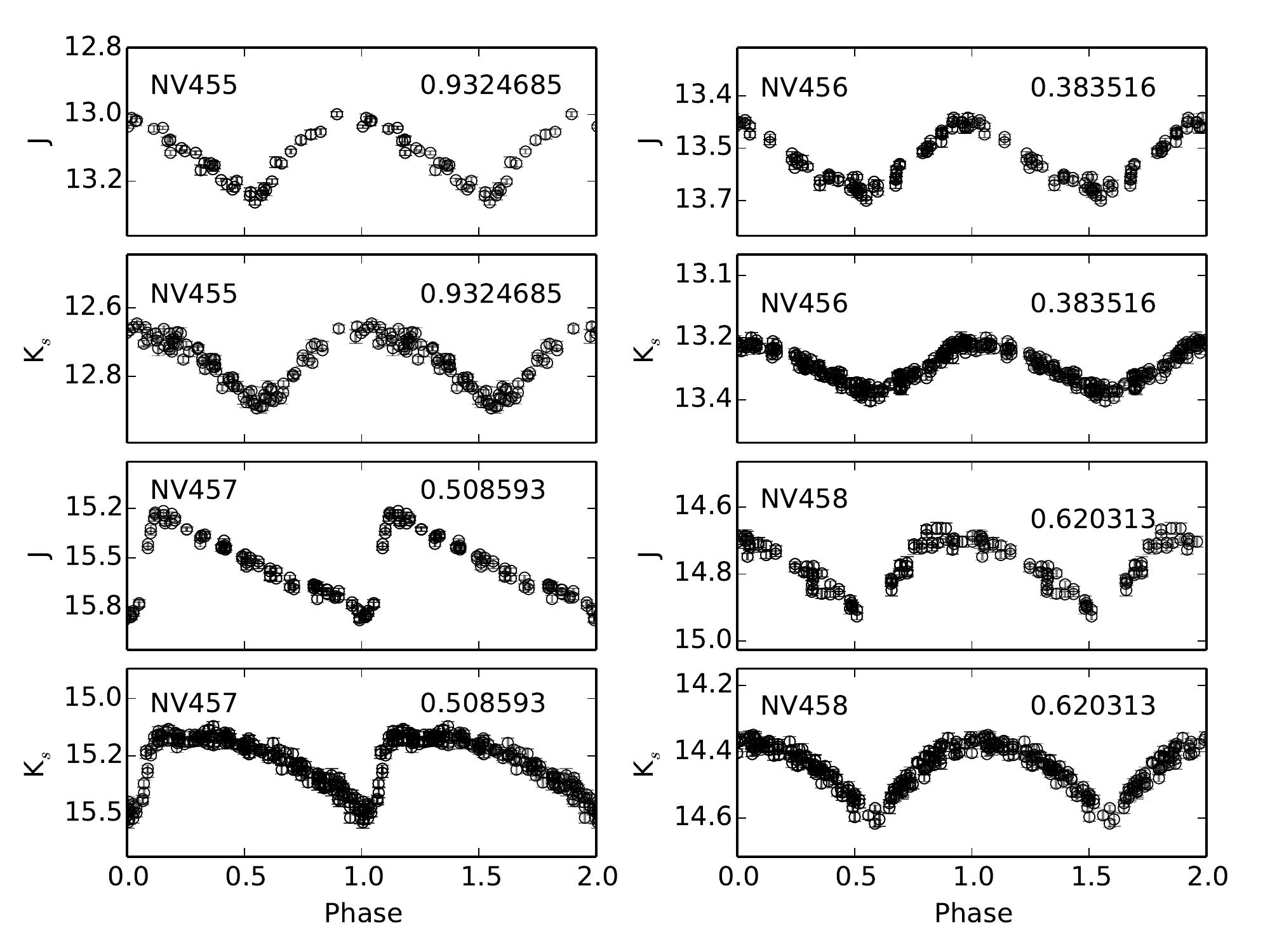}
   \caption{$J$ and $K_{\rm S}$-band light curves of the newly discovered RRL stars that are likely
   cluster members.}
   \label{lcurvesmf}
   \end{center}
   \end{figure*}

  Table~\ref{table:2} gives the ID (following the numbering scheme from the C13 online 
  catalogue), coordinates, intensity-weighted mean $J$ and $K_{\rm S}$ magnitudes, 
  periods, type, and the distance to the center of the cluster. 
  According to their position in the CMD, NV455 and NV456 
  are likely members of the cluster, whereas the other two stars deviate largely from the main HB 
  locus~-- and since they are fainter, they are almost certainly field stars in the cluster background. 
  $J$- and $K_{\rm S}$-band light curves are presented in Figure~\ref{lcurvesmf}. 
  According to their light curve shapes, 
  in addition to their periods (Table~\ref{table:2}) and positions in the Bailey diagram (Fig.~\ref{bailey}), 
  we assign an RRc type to the new variable NV456, and an RRab type to variables NV455, NV457, 
  and NV458. As it happens, NV455, with a period of about 0.932~d, is the RRab star with the third 
  longest period known to date in $\omega$~Cen (see Fig.~\ref{bailey}).
   
  Recently, \cite{F13} also performed a variability search in the $V$-band over an area of 50~deg$^2$
  around $\omega$~Cen, with the main aim to detect RRL stars and in this way find evidence for tidal 
  debris from the putative dwarf galaxy progenitor of the cluster \citep[e.g.,][]{dd02,maea05,C10,vc11}. 
  They found 29 new RRL stars, and recovered 7 previously known ones. Of these, a total of 11 are 
  in the area covered by our observations, and they were all independently recovered in our study as 
  well. Their remaining RRL are all outside $\omega$~Cen's tidal radius, and they are reported further 
  on in \cite{FT14}.
  
  \section{Summary}\label{conclusions}
  
  Based on previous $\omega$~Cen variability studies, from \cite{B02} until \cite{K04}, 
  C13 report that there are 193 RRL stars in the field of the cluster. Among these, one 
  finds 92 RRab, 100 RRc, and 1 star classified as RR? (V80). None of the 5 new RRL claimed  
  by \cite{W07} were included in the C13, catalogue, because, as we have seen, only two of 
  them are really new discovered~-- and for these two stars, ID-91 and ID-99, an RRL 
  classification still remains uncertain. 
  
  On the other hand, ID-74, which \cite{W07} classified simply as ``pulsating,'' was included in 
  the C13 catalogue as a possible RRL, based on its period (0.6671~d). In this way, it was also 
  given the official ID V433 by C13. However, we have been unable to recover the star's variability 
  using our NIR data. Moreover, its NIR magnitude and color place the star close to but to 
  the red of the base of the red giant branch, rather than on the classical IS. Recently, \cite{N13} 
  found a misidentified RRab star, V411, in the \citeauthor{W07} catalogue. Therefore, prior to 
  this publication, 194 RRL stars in all (93 and 100 RRab and RRc, respectively) have been listed 
  for the $\omega$~Cen field, where we have not included any of the ``new'' candidates in the 
  \citeauthor{W07} catalogue in this number count. 

  Considering the information obtained with the VISTA observations, the light curves of some RRLs 
  were recovered for the first time, leading to changes in the variability classification. In 
  particular, we have been able to provide the first light curves ever for V80, thus establishing
  that it is an RRc star. V151, V173, and V177, all of which had previously been considered ab-types, 
  are shown to be RRc stars, according to our data. On the other hand, we have found that V180 is not 
  an RRc, as previously thought, but rather a contact binary of the W~UMa type. Furthermore, 4 new 
  variables were found using VISTA: 3 RRab and 1 RRc stars. Therefore, there are now 197 known RRLs 
  in the $\omega$~Cen field, including 93 RRab and 104 RRc stars. 
  
  From these 197 RRLs, there are 4 RRab (V181, V283, NV457, and NV458) and 2 RRc (V168 and V183) 
  stars that have magnitudes and colors consistent with being field interlopers. In addition, if 
  we exclude V68 and V84 because of their controversial classification (with the distinct 
  possibility that they may be ACEP stars, rather than RRLs), the final census of likely RRL 
  members of $\omega$~Cen includes 88 RRab and 101 RRc stars, for a grand total of 189. 
  
  It should be noted that it was not possible for our team to observe some distant RRLs such as 
  V171, V178, V179, and V182, because they fall outside the VIRCAM@VISTA FoV; however, they are 
  located outside the tidal radius of the cluster, and are therefore most probably field stars. 
  It is expected that these numbers will increase further, once the new variables found by 
  \cite{F13} in the $\omega$~Cen neighborhood are published.  
  
  In a forthcoming paper, we will present a detailed analysis of the NIR 
  period-luminosity relations followed by RRLs as well as type II Cepheids and SX Phoenicis 
  stars in the $\omega$~Cen field, based on our VISTA observations. This will be followed by 
  a paper providing the complete, updated catalogue of variable stars in $\omega$~Cen, including 
  the full list of RRL, type II Cepheids, SX~Phoenicis stars, eclipsing binaries, and 
  anomalous Cepheids, along with their NIR properties.

\begin{acknowledgements}
      We warmly thank A. K. Vivas for useful discussions and information regarding her team's 
      variability search around $\omega$~Cen. We also thank the anonymous referee for   
	  useful comments that helped improve the presentation of our results. 
      Support for this project is provided by the Ministry for the Economy, Development, and 
      Tourism's Programa Iniciativa Cient\'{i}fica Milenio through grant IC\,120009, awarded to the
      Millennium Institute of Astrophysics (MAS); by Proyecto Basal PFB-06/2007; and by FONDECYT 
      grants \#1141141 (C.N., F.G., M.C.), \#1130196 (F.G., D.M.), \#3130320 (R.C.R.), and 
      \#3130552 (J.A.-G.). C.N. and F.G acknowledge additional support from CONICYT-PCHA/Mag{\'i}ster 
      Nacional (grants 2012-22121934 and 2014-22141509, respectively). 
\end{acknowledgements}

\bibliography{adssample}

\appendix

  \section{Notes on individual RR~Lyrae stars}\label{Notes}
 
  \textbf{\textit{V19 and V21}}:~ In the C13 catalogue, these variables are listed as possible 
  non-members, based on their anomalous Fourier parameters compared with other variables with 
  similar periods \citep{C00}. However, and as indicated in Table~\ref{table:1}, the proper 
  motion studies of vL00 and B09 found a membership probability $\mathcal{P}_{\mu}$ of (100\%,
  \,100\%) and (100\%,\,99\%) for V19 and V21, respectively. The membership status for these 
  two stars is further supported by their position in our NIR CMD, which locates them in 
  the middle of the RRL IS (see the bottom panel of Fig.~\ref{CMD}).

  \textbf{\textit{V52}}:~ This star appears to be much brighter than other RRab stars in $\omega$~Cen, 
  as can be noted in Figure~\ref{CMD}. Its light curve has high photometric errors despite its 
  high luminosity, thus suggesting contamination due to blending. Indeed, vL00 reported 
  that V52 may have an unresolved companion, which would account for the fact that it is brighter 
  and has a lower amplitude than other RRL variables with the same period. According to vL00 
  and B09, the membership probability of this star is 42\% and 53\%, respectively. Based on its  
  position in our NIR CMD, the projected distance to the cluster center, $d \sim 2 \arcmin$, 
  and the high amount of noise in the light curve, V52 is most likely a member of the cluster in a 
  very marked blend. Inspection of {\em Hubble Space Telescope} ({\em HST}) images \citep{jaea10} 
  confirms the presence of a companion that is unresolved in our images, located just about 0.5\arcsec 
  from the RRL star, but brighter by about 0.5~mag in $V$, and bluer by about 0.25~mag in $B-V$.   

  \textbf{\textit{V56 and V168}}:~ The field status of these two stars was based on a radial 
  velocity study by \cite{L81}. Subsequently, V56 was again considered a member of the cluster 
  based on its high membership probability of 98\% and 100\% found by vL00 and B09, respectively~-- 
  a status in excellent agreement with its position in our NIR CMD. These inconsistent 
  conclusions can probably be explained as follows. 
  When assessing membership status of an RRL variable from its radial velocity, it is necessary 
  to take into account the phase in its light and velocity cycle. \citeauthor{L81} obtained this 
  information from R. J. Dickens (1980, priv. comm.), based on an epoch established in 1974. The 
  observations of \citeauthor{L81} were made in 1978. A period change study by \citet{jjea01} 
  indicates that V56 undergoes random period changes. Thus it is highly likely that the light 
  curve phase assumed by \citeauthor{L81}  was incorrect. On the contrary, all proper 
  motion studies agree in classifying V168 as a non cluster member, which also agrees with our 
  photometric classification as a field star: Figure~\ref{CMD} clearly shows that this RRL star 
  is a background interloper.

  \textbf{\textit{V68 and V84}}:~ These two stars have controversial mode classification in 
  the literature. First thought as RRLs by \cite{B02}, they were later suggested to belong to the 
  ACEP group by \cite{N94}, because of its brighter optical magnitudes compared 
  to the other RRLs of the cluster. ACEP stars are Population II stars with periods between 0.4 
  and 2.4~days, that do not follow either the RRL or the type II Cepheid PL relations. Most of 
  them are found in dwarf spheroidal galaxies \citep{N94} and the Large Magellanic Cloud \citep{S08}, 
  a few ACEPs are known in Galactic globular clusters \citep[see ][]{Z76, S10, O12, A13} and only 
  one in the Galactic field \citep[XZ~Ceti;][]{T85, S07}.
  
  V68 appears as an RRc cluster member star according to its proper motions and its position in 
  the CMD (see Table~\ref{table:1} and the bottom panel of Fig.~\ref{CMD}). Although it has the longest 
  period among the RRc stars of the cluster (0.56~days), which is longer than even the short-period end  
  of the RRab stars' period distribution, 
  the amplitude of its light curve is small enough to place it within the RRc 
  stars group (see Fig.~\ref{bailey}). V68 also follows the PL relations for RRc stars in $J$ and 
  $K_{\rm S}$ (Navarrete et al. 2014b, in preparation). This star, on the other hand, also appears  
  to follow the optical ($B$, $V$) PL relations for ACEP stars, as derived by \cite{N94}.
  
  On the other hand, V84 was discarded as a cluster member by vL00, with a membership probability  
  $\mathcal{P}_{\mu} = 0\%$. Unfortunately, we were unable to confirm this result using proper motions 
  from B09, since V84 
  is outside their FoV. Our NIR CMD shows that V84 is well located in the RRL instability 
  strip, opening the possibility that it is a cluster member. However, as \cite{DP06} pointed out, 
  V84 does not follow the NIR PL relation for RRab stars. This result is confirmed using our  
  PL relations (Navarrete et al. 2014b, in preparation), with V84 appearing brighter than other stars with 
  the same period. Conversely, \cite{N94} found that V84 was placed slightly below the $B$- and $V$-band 
  PL relations for ACEPs, which could be explained if the star was not an ACEP but instead an RRL star 
  leaving the HB.
  
  Using our NIR observations, we cannot adopt a definitive classification for either of these stars. 
  If they are RRLs, the most probable scenario is that V68 is an RRc member and V84 is an RRab 
  star from the field. However, the ACEP classification cannot be completely discarded. Indeed, 
  V68 and V84 follow the 
  $K_{\rm S}$-band PL relation for ACEPs in the Large Magellanic Cloud derived by \cite{R14},
  assuming a distance modulus for $\omega$~Cen of 
  $(m-M)_0 = 13.63 \pm 0.1$ mag~-- which is in good agreement with the one derived by 
  \cite{M06} using type II Cepheids. If these two stars are indeed ACEPs, they should be members 
  of the cluster.
  
\begin{figure*}
   \centering
   \includegraphics[width=0.8\textwidth]{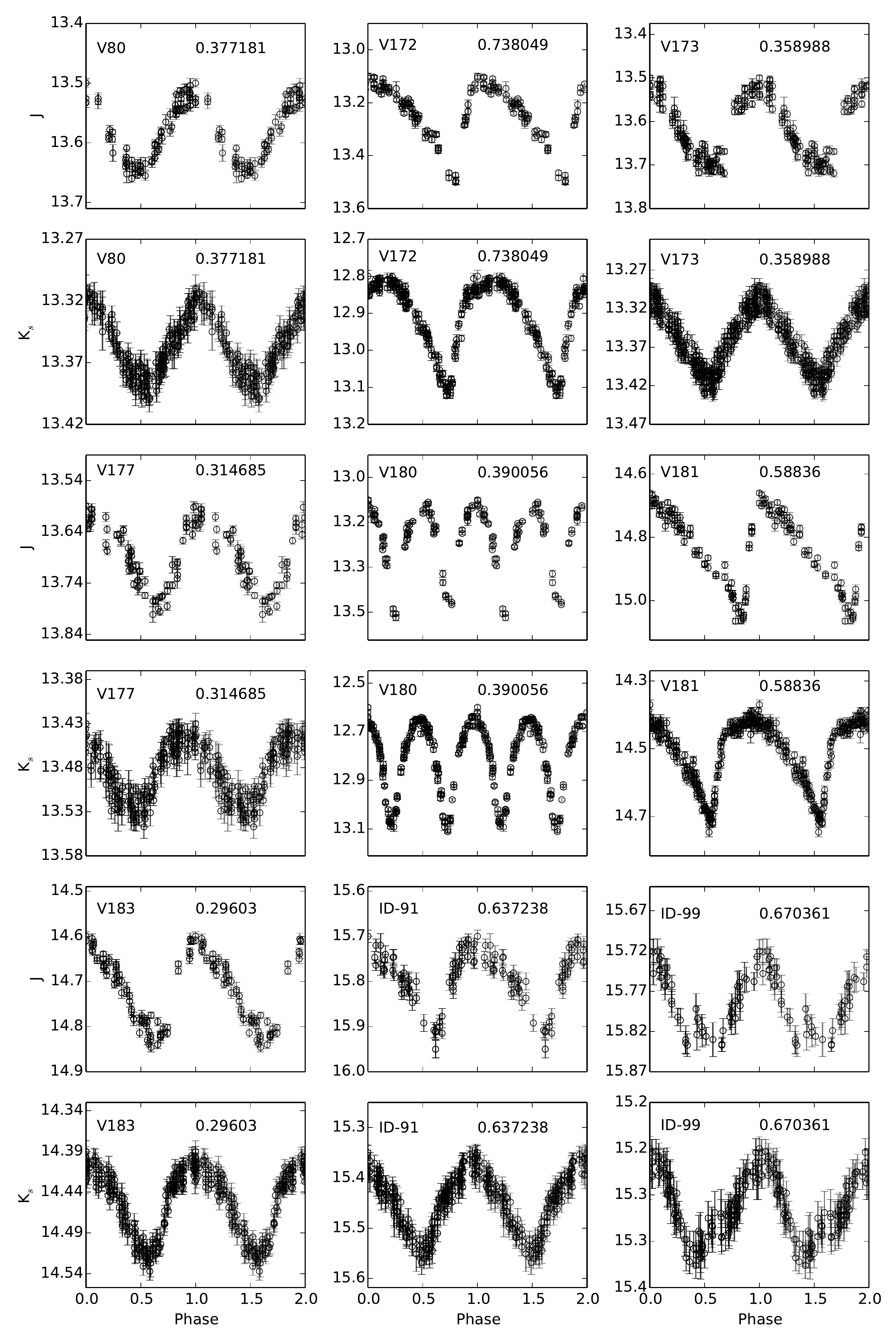}
   \caption{VISTA NIR light curves for all the RRLs that lacked a derived light curve or 
   period in the literature. For each star, the upper panel shows the $J$-band curve, whereas the 
   lower one displays the $K_{\rm S}$ band instead. Our derived period for each star is provided 
   in each panel. Note that the V180 light curve is consistent with a W~UMa classification. 
   Variables ID-91 and (especially) ID-99, from \cite{W07}, 
   have light curve shapes that are different from all the other 
   RRab in $\omega$~Cen, and are suspected field stars.
   }
   \label{lcurves}
   \end{figure*}

  \textbf{\textit{V80}}:~ This star was detected in the first variability study on $\omega$~Cen carried 
  out by \cite{B02}. In that work, V80 was reported as difficult to measure because it is distant 
  from comparison stars, and accordingly only sparse measurements of differential magnitudes were published, 
  without an accompanying light curve. \citeauthor{B02} suggested a period of 0.45 or 0.31~days for this star, 
  wherewith it was tentatively classified as an RRL-type star. Based on our wide-field VISTA 
  observations, we determined its period, $P = 0.37718$~d (Table~\ref{table:1}), and derived  
  its complete, phase-folded light curve (Fig.~\ref{lcurves}). 
  
  According to its coordinates, ID-53, $P = 0.377$~d, from \cite{W07} appears as the nearest star 
  to V80 among the list of new variables discovered by those authors. However, \citeauthor{W07} did 
  not associated ID-53 to V80, probably because a period value had never previously been published 
  for V80~-- and accordingly, they claimed that their ID-53 was a newly discovered variable star. 
  Instead, our independently derived period and coordinates alike indicate that V80 and ID-53 are 
  one and the same star. 
  
  V80 is not present in any proper motion study. Its derived period and position in the CMD 
  indicates that V80 is an RRc star that is also a member of the cluster.

  \textbf{\textit{V118, V135, and V139}}:~ These three stars appear well located within the RRab 
  region in the IS (see Fig.~\ref{CMD}). Based on the vL00 and B09 proper motion studies, V118 has 
  a membership probability of $\mathcal{P}_{\mu} =$~(100\%, 96\%), respectively, whereas V139 has 
  $\mathcal{P}_{\mu} =$~(100\%, 90\%). V135 is only present in B09, with a membership probability 
  of 97\%. However, all appear brighter than the other RRab stars with similar periods in the $J$ 
  and $K_{\rm S}$ PL relations (Navarrete et al. 2014b, in prep.).

  V118 is located in a very crowded region. According to the {\em HST} catalogue of \cite{jaea10}, 
  there are more than 10 stars inside a $1\arcsec$ radius around its coordinates. A similar 
  situation occurs with V139, having a neighbor located at only 0.432\arcsec, according to the 
  \citeauthor{jaea10} catalogue~-- which is a difference of less than 2~pixels, given the VIRCAM 
  pixel scale. Hence, this star is almost certainly brighter in our photometry than it would be, 
  in case it were isolated.

  V135 is suffering the influence of a disturbing neighbor, located $2.5\arcsec$ away, which is 
  saturated. This situation was previously reported by \cite{S09}, with both stars being labeled 
  in that study as a visual binary.
  
  \textbf{\textit{V143}}:~ It is certainly a member of the cluster according to vL00, with 99\% 
  membership probability. B09 give instead $\mathcal{P}_{\mu} = 90\%$ for this star. However, it 
  appears a few tenths of a magnitude   brighter than other RRab stars with the same color. 
  A visual inspection of this star in some of our images suggested to us that its anomalous 
  position in the CMD is not due to a blend effect of an unresolved system, but rather to a 
  saturated companion located close by (i.e., at a distance of $\approx 3.4\arcsec$) that is 
  somehow disturbing the photometric quality in its neighborhood. Indeed, this situation was 
  explicitly reported by \cite{S96}, where the presence of a bright star located southwest from 
  the position of V143 is remarked upon. Therefore, despite its anomalous position in the CMD, 
  we suggest that V143 is indeed a cluster member.
  
  \textbf{\textit{V151}}:~ Listed as an RRab star by \cite{K04} with a period of 0.4078~days, this 
  star would appear to be the shortest-period fundamental-mode RRL in $\omega$~Cen. However, 
  despite the fact that our estimated period ($P = 0.407756$~d) is in excellent agreement with the 
  one derived by \cite{M38}, we notice that its light curve shows a very small amplitude, which  
  is among the lowest for RRab stars in our study. Small amplitudes are unusual for short-period 
  RRab stars. Figure~\ref{bailey} shows the position of V151 in the Bailey (period-amplitude) 
  diagram, as compared to other $\omega$~Cen RRL stars. This plot clearly shows that V151 is well 
  located among the RRc-type variables. In addition, its position in the PL relation (Navarrete et 
  al. 2014b, in preparation) is also consistent with first-overtone pulsation. Thus, we conclude 
  that this star should be classified as an RRc. We note that \cite{M38} also classified V151 as 
  an RRc star, hence the erroneous RRab type adopted by \citeauthor{K04} is likely a typo, which 
  was later inadvertently reproduced in the C13 online catalogue. V151 was not included in the 
  proper motion studies of vL00 and B09; however, our NIR CMD suggests that this variable 
  likely belongs to the cluster.

  \textbf{\textit{V172, V173, V175, V177, and V180}}:~ \cite{W65} announced the discovery of V171-V180. 
  Of these 10 stars, V176 was subsequently discarded as a non-variable star, V174 classified as an EB, 
  and the remaining stars classified as RRab's, except for V180, which was classified as an RRc instead. 
  To the best of our knowledge, no further variability studies of these stars have been carried out, 
  except for V172 and V177 which were included in the observations by \cite{W07} as ID-18 and ID-49, 
  respectively. None of them has had a period assigned in the C13 catalogue.  

  Four of Wilkens's stars, namely, V171, V175 (without a previous type assigned), V178, and V179 are 
  all very far from the cluster center, and thus were not observed by us, despite VIRCAM's large FoV. 
  Consequently, only light curves for V172, V173, V177, and V180 were derived, and they are shown in 
  Figure~\ref{lcurves}. Only V180 has membership probability derived based on proper motions studies, 
  being considered a non-member of the cluster with a null membership probability by vL00. According 
  to their positions in the CMD (bottom panel of Fig.~\ref{CMD}), we suggest that V172, V173, and 
  V177 are cluster members. We confirm the ab-type classification for V172, but at odds with Wilkens, 
  we did not find an RRab behavior for either V173 or V177, and we suggest instead an RRc classification 
  for both these stars. This is consistent with their positions in the Bailey diagram, as can be seen 
  from Figure~\ref{bailey}. V175, according to A. K. Vivas (2014, priv. comm.; see also \citealt{F13}) 
  is a foreground c-type RRL. As can be seen in Figure~\ref{CMD}, V180 is located in the RRab region, 
  redder than any other RRc star. Moreover, it has a large amplitude compared to other RRc stars 
  (Fig.~\ref{bailey}). Our estimated period produces a light curve clearly showing 2 different 
  minima. Accordingly, we propose to classify V180 as a contact EB of the W~Ursae Majoris type, 
  rather than as an RRc, which is consistent with its fairly red color and the fact that this star 
  is not a cluster member. 

  \textbf{\textit{V181 and V183}}:~ V181, V182, and V183 were discovered by Wesselink (1969, 
  unpublished, priv. comm. to H. Sawyer-Hogg), who indicated the stars' coordinates and 
  periods, but without providing light curves or variability types for these stars. \cite{C01}, 
  based on Wesselink's periods, classified V181 and V182 as RRab's, and V183 as an RRc. 
  These stars were outside the FoV in subsequent investigations by other authors. Unfortunately, 
  V182, at 59.7~arcmin from the center of the cluster, is also outside the FoV of VISTA. 
  
  Figure~\ref{lcurves} shows the first complete light curves for V181 and V183. They are ab- and 
  c-type RRL stars, respectively, in agreement with the periods originally derived by Wesselink.  
  These two stars are very distant from the cluster center (though still inside its tidal radius), 
  and thus they are absent from proper motion studies of the cluster. However, based on their 
  positions in the CMD (see the top panel of Fig.~\ref{CMD}), they both appear to be field RRL stars.

  \textbf{\textit{V268}}:~ vL00 claimed that V268 is a field star with a membership probability of 0\%. 
  However, we found some inconsistencies in this membership status using our coordinates and those 
  from the literature: in fact, the nearest star to V268 in vL00's catalogue has a membership probability 
  of 70\%. Moreover according to B09, this star has a 99\% membership probability. Using the NIR 
  CMD to figure out the membership status of this star, it is found that V268 is located well 
  inside the RRL instability strip, as the other RRab members of the cluster (Fig.~\ref{CMD}, bottom panel). 
  We thus conclude that V268 is indeed a cluster member. 
  
  \textbf{\textit{V283}}:~ This star was studied by neither vL00 or B09. Based on its faint 
  visible magnitude, \cite{K97} claimed that it is a background halo RRL. This field 
  status is confirmed by our photometry, which locates the star a few magnitudes below the 
  HB, even fainter than the turn-off point level (see CMD in the top panel of Fig.~\ref{CMD}).
  
  \textbf{\textit{V349 and V351}}:~ These two RRc variables were discovered by \cite{K04} and are 
  located $\sim 1.5$~arcsec away from the cluster center. The huge amount of crowding at their 
  positions did not allow those authors to present properly calibrated light curves. 
  Interestingly, \citeauthor{K04} noted that V351 is a multiperiodic variable, and suggested 
  that it may show non-radial pulsation modes, as previously proposed by \citet{C00} for several 
  other RRL stars in $\omega$~Cen. 
  Using the coordinates and periods provided by \cite{K04}, these two variables were not found neither  
  in our ALLFRAME nor in our DoPhot photometry. The matching radius used was 1~arcsec. Increasing the 
  matching radius until 10~arcsec still did not return any variable stars. The last check step was to 
  consider all the stars within 30~arcsec around the listed coordinates, and check for the presence 
  of periodicity in this enlarged sample using the Analysis of Variance statistic \citep[ANOVA;][]{S89}. 
  Again, no periodic variables were detected in this way. Therefore, V349 and V351 
  could not be recovered using PSF photometry in our VISTA images.

  \section{RR~Lyrae stars from \cite{W07}}\label{Weldrake}
  
  \cite{W07} reported the discovery of five new RRL stars in $\omega$~Cen: ID-53, ID-91, ID-99, 
  ID-144, and ID-145. By carefully checking these stars' positions and periods, we have been 
  able to establish that three out of these five variables were not really new discoveries. 
  More specifically, according to our analysis, we were able to establish that ID-53~= V80, 
  ID-144~= V272, and ID-145~= V15. 

  ID-91 and ID-99, on the other hand, do appear to be bona-fide new variables. \cite{W07}   
  listed optical magnitudes of $V\sim 17.1$ and $17.45$~mag for these two stars, respectively, 
  which is consistent with their being field interlopers. Unfortunately, neither star appears
  in the proper motion catalogues of vL00 and B09. As far as their variability status is
  concerned, \citeauthor{W07} suggested that 
  ID-99 could be either a long-period RRc with a 0.671-d period or an EB with a period twice 
  as long. For ID-91, in turn, they assigned an RRab type. Both stars' light curves are shown 
  in Figure~\ref{lcurves}, according to which the RRab status of ID-91 seems reasonable, 
  whereas the RRL nature of ID-99 remains a possibility. However, if indeed an RRL, ID-99's 
  long period might more naturally suggest an RRab classification, as opposed to the RRc type 
  suggested by \citeauthor{W07}. 
  To further check their classification, ID-91 and ID-99 were plotted in the Bailey diagram 
  (see Fig.~\ref{bailey}). Their location suggest that both stars are either extreme examples of 
  OoI RRab stars, and/or show a modulation in their light curves (e.g., the Blazhko effect), 
  and/or are unresolved blends. The latter possibility receives some support 
  from the positions of the stars in the CMD (Fig.~\ref{CMD}), which show that both stars are  
  slightly redder than expected for RRL stars. However, these stars 
  are also quite far away ($\sim 25\arcmin$) from the cluster center, and inspection of our images reveals that 
  they are in relatively unpopulated areas, thus making the possibility of blends seem unlikely. 
  Unfortunately, {\em HST} images of these fields are not available, and thus we cannot 
  conclusively establish that unresolved companions to these stars are indeed absent. 
  In like vein, the presence of the Blazhko effect is not supported by either our photometry 
  (Fig.~\ref{lcurves}) or the \cite{W07} photometry. 
  
  Based on the astrometry that is provided along with the VISTA images (which is tied to the 2MASS system),  
  we found some erroneous matches between some of the 
  known variables in the catalogues of \cite{W07} and \cite{K04}: ID-115~= V264 (instead of V48) and 
  ID-135~= V266 (not V356). Moreover, \citeauthor{W07} erroneously claimed that ID-133 was the same star 
  as V144 from the \cite{C01} catalogue. However, as already reported by \cite{N13}, ID-133 is indeed 
  a different variable star. C13 has already added this new RRab star in her online catalogue, where 
  it is now officially listed as V411. 
  
  C13 has included ID-74 of \cite{W07}, a possible RRL based on its period (0.6671~d, according to 
  \citeauthor{W07}), 
  as V433 in her catalogue. This star has 
  $V \approx 16.5$ and $(V-I)\approx 1.0$~mag, which suggests that it is a field star. Indeed, according 
  to our NIR magnitudes, V433 star is placed above the cluster's turn-off point, above the main 
  sequence and close to (but redder than) the base of the red giant branch (see the top panel of Fig.\ref{CMD}).
  Still, the star does appear to show some 
  short-period variability, with a period of $(0.6681 \pm 0.0004)$~d,
  and an amplitude that is likely not higher than 0.05~mag in $K_{\rm S}$, 
  again preventing us from classifying it as an RRL. The star's red color is unlikely to be caused 
  by a blend with a red star, since it is $\sim 30\arcmin$ away from the cluster center, and without 
  any obvious close companions in our images. According to \citet{dcc08}, the star is a radial 
  velocity non-member. 
  In conclusion, we follow \cite{W07}, and consider 
  V433 as a variable star with an uncertain variability type.

\end{document}